\documentstyle[prl,aps,epsf,multicol]{revtex}
\begin{document}
\tighten

\title{Entropically driven reentrant SmC-SmA-SmC phase transition
in composite polymer--liquid crystal systems}
\author{Leo Radzihovsky}
\address{Department of Physics, University of Colorado, Boulder, 
CO 80309\ }
\date{\today}
\maketitle
\begin{abstract}
I consider the effects of polymers on the smectic phase of a host
liquid crystal matrix.  Focusing on the regime in which the polymers
are predominately confined between the smectic layers, I find that the
presence of the polymers can lead to a reentrant phase diagram with
the smectic-C sandwiching the smectic-A phase from both the high and
low temperature sides. Simple entropy-energy arguments predict the
shape of the reentrant phase boundary.
\end{abstract}

\begin{multicols}{2}
\narrowtext

Recently, some of the attention in liquid crystal research has
focused on composite polymer--liquid crystal systems, with the aim
of expanding the range of applications of liquid crystals by
controlling the electro-optical properties and phase behavior. 

This theoretical interest in such systems is motivated by recent
experiments in which a two-component smectic liquid crystal--monomer
mixture is found to segregate into monomer--rich layers confined
between the smectic-A (SmA) layers\cite{Colorado}. In these experiments a
significant increase in polymerization rate was observed when
polymerization was initiated within the smectic phase of the liquid
crystal host. The confining smectic layers are thought to lead to a
two-dimensional (2d) organization of monomers between the layers, thereby
enhancing the polymerization rate. It is believed that the final
polymerized state is dominated by configurations in which polymers are
confined between the smectic layers. Similar segregation of organic
solvents intercalated between the smectic layers of a host
thermotropic smectic liquid crystal was observed in recent X-ray
experiments by Rieker\cite{Rieker}.

\begin{figure}[tbh]
\centering
\setlength{\unitlength}{1mm}
\begin{picture}(70,60)(0,0)
\put(-30,-70){\begin{picture}(70,60)(0,0) 
\includegraphics{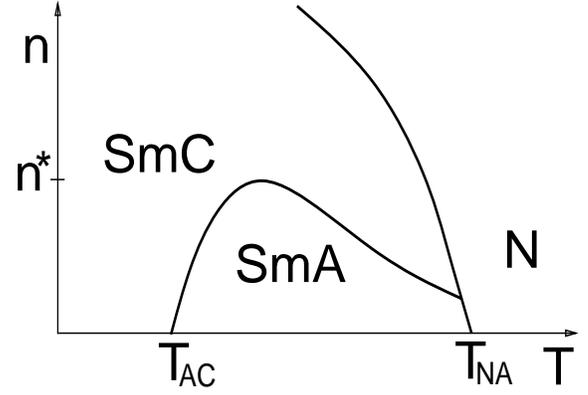} 
\end{picture}}
\end{picture}
\caption{The proposed phase diagram for the smectic liquid crystal
with polymers confined between the smectic layers as a function of
temperature $T$ and the polymer density $n$.} 
\label{fig1}
\end{figure}

In this paper I examine how the presence of the polymers confined
between the smectic layers modifies the SmA and SmC regions
of the phase diagram. I find that polymer presence can lead to the
entropically induced reentrant SmC-SmA-SmC phase diagram displayed in
Fig.\ref{fig1}.

The standard expression for the free energy of the
smectic phase which allows for the possibility of developing the
SmC order parameter ${\vec C}$, describing a 2d
projection of the nematogens onto the smectic layers is
\begin{eqnarray}
F_{lc}&=&\int\bigg[{K_1\over2}(\nabla_\perp^2 u)^2 +
{B\over2}(\partial_z u)^2+
\alpha(\vec{\nabla}_\perp\cdot\vec{C})(\nabla_\perp^2 u)\nonumber\\
&+&{J\over2}|\nabla\vec{C}|^2 + 
{1\over2}(T-T_{AC})|\vec{C}|^2 +
{u\over4}|\vec{C}|^4\bigg]\;.
\label{Flc}
\end{eqnarray}
The first two terms describe the smectic elasticity. The third term
describes the energetic tendency of the local inhomogeneity in the
tilt order parameter to induce the local layer extrinsic curvature,
which screens the nonuniformity in $\vec C$. The last three terms are
the contribution due the SmC tilt order, which develops for $T<T_{AC}$
via a well studied continuous phase transition in the 3d XY
universality class that has been extensively studied and is quite well
understood\cite{deGennes}.  Within the SmC phase, the order parameter
$\vec C$ can be integrated out and the resulting elastic free energy
is similar to the SmA free energy except for the in-plane 
anisotropy, generated by the average molecular tilt.

How is the well-studied Nematic-SmA-SmC phase diagram modified by the
presence of polymers?  Obviously the smectic will swell in the
$z$-direction, leading to a $T$-- and polymer density ($n$)--dependent
liquid crystal density modulation wave vector $q_0(T,n)=2\pi/d(T,n)$.
The layer spacing $d(T,n)$ will increase with $T$ and $n$, variation
of which should be observable in X-ray scattering
experiments\cite{Rieker}. The dependence of $d(T,n)$ on $n$ can be
simply evaluated\cite{Rieker} by assuming for simplicity that large
fraction of added polymer is concentrated {\em between} smectic layers
(an assumption consistent with the experimental
observations)\cite{Colorado}. Assuming pure SmA order I estimate
$d=d_0+\Delta d_0=d_0(1+f_v)$, where $f_v=V_p/V_{lc}$ is the
polymer-to-liquid crystal volume fraction. The $T$-dependence can be
obtained from the estimate of the entropic polymer pressure (see below).

\begin{figure}[bth]
\centering
\setlength{\unitlength}{1mm}
\begin{picture}(150,50)(0,0)
\put(-20,-88){\begin{picture}(150,0)(0,0) 
\includegraphics{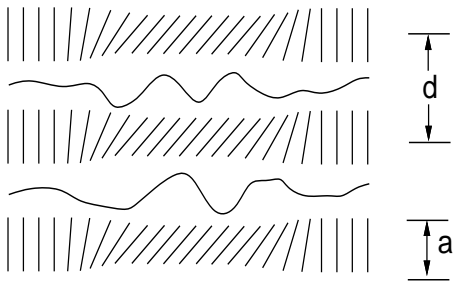} 
\end{picture}}
\end{picture}
\caption{Schematic illustration of how polymers confined between the
smectic layers generate an effective entropic interaction that induces
local SmC tilt order: the regions with finite tilt order allow for
larger out-of-plane polymer undulations, thereby lowering the entropic
free energy.}
\label{fig2}
\end{figure}

A much more interesting consequence (that is likely to dominate in a
confined geometry of e.g. a liquid crystal display) is that polymers,
confined by the smectic layers, will force the nematogens to tilt
relative to the smectic layer normal. This tilt will increase the {\em
effective} space available for polymer diffusion between the confining
layers, thereby lowering the entropic polymer pressure (see
Fig.\ref{fig2}). Therefore such entropic polymer-smectic interaction
will induce a local SmC order parameter $\vec C$, which can develop
into a true long-range order.

The reentrant behavior can therefore be understood in the following
way. For $T>>T_{AC}$, the increase in the total free energy due the
confinement of polymers is large and can be accommodated by developing
SmC order, thereby stabilizing SmC phase over the SmA phase. In the
intermediate $T$ range, however, the polymers' tendency for diffusion
and therefore the entropic interaction is suppressed and the SmA phase
is again stabilized. Finally, for $T<T_{AC}$ the SmC is stabilized
over the SmA phase, as in the polymer-free liquid crystal.

To support above physical arguments I calculate the
entropically-induced polymer interaction $\delta F_p$ for the SmC
order parameter. I take the center of mass layer spacing to be
$d=d_0+\Delta d$, with $\Delta d$ to be determined below, and consider
the decrease in polymer entropy due to the confinement of spacing
$Z_o$ in the $z$-direction (related to $|\vec C|$).

In the absence of confinement a polymer of length $L$ would explore a
3d region of space whose extent in the $z$-direction is in general
$R_G\sim{l_p(T)} (L/l_p(T))^{\nu_z}$, where the wandering exponent
$\nu_z$ and the persistent length $l_p(T)$ are determined by polymer
interaction, and by the relevant range of $L$ related to the smectic
spacing $d$ and polymer elasticity (see below).  For the smectic
confinement of width $Z_o$, a polymer will typically make collisions
with the smectic layers every $L_o\sim
l_p(T)^{1-1/\nu_z}Z_o^{1/\nu_z}$ of internal length. In the total
length $L$ of a single polymer these excursions will on average result
in $N_c\sim L/L_o\sim L l_p(T)^{1/\nu_z-1}Z_o^{-1/\nu_z}$ collisions
with the smectic layers. Since upon each of these (predominantly)
reflecting collision the polymer is prevented from continuing along
the $z$-direction, each encounter with the smectic layer leads to
entropy {\em reduction} by $s\approx\log2$. The corresponding total
{\em increase} in the free energy density is given by
\begin{equation}
\delta F_p\sim T n s N_c\sim\left({T l_p(T)^{1/\nu_z-1}\over
Z_o^{1/\nu_z}}\right) n s L\;,
\label{dF}
\end{equation}
where $n$ is the 3d density of polymers. 

As can be seen from Fig.\ref{fig2} the SmC order parameter is directly
related to the maximum allowed excursion $Z_o(|\vec{C}|)= d -
\sqrt{a^2 - |\vec {C}|^2}$, where $a$ is the length of the liquid
crystal molecule. Combining this expression with Eq.\ref{dF}, assuming
for simplicity that $|\vec C|^2\ll a < d_0$ and $\Delta d\ll d_0$, and
expanding in these small quantities, I find
\begin{equation}
\delta F_p\approx\mbox{const}-{1\over2}T
l_p(T)^{1/\nu_z-1}\tilde{n}\big(|\vec C|^2+2a\Delta d\big)\;,
\label{dFii}
\end{equation}
where $\tilde{n}=(2 n s L)/\left(\nu_z a(d_0-a)^{(2/\nu_z+1)}\right)$ 
and obviously from Fig.\ref{fig2} $d_0>a$. 

As argued in the introduction and expected on physical grounds, this
entropically induced part of the free energy can be reduced by
swelling the layers ($d$) $\Delta d>0$ and/or by inducing the SmC
order $|\vec C|>0$. The value of the swelling $\Delta d$ can be
roughly determined by balancing the above negative free energy against
the positive elastic strain energy $E_e\sim (B/2)(\Delta d/d_0)^2$, ($B$ is
the compressional bulk modulus), giving 
\begin{equation}
\Delta d(T)\approx T l_p(T)^{1/\nu_z-1}\tilde{n}a d_0^2/B\;,
\label{Delta_d}
\end{equation}
This increase in the layer spacing $d$ leads to a 
($\vec C$-independent) reduction of the above free energy by an
amount $(T l_p(T)^{1/\nu_z-1}\tilde{n}a d_0)^2/(2B)$.

To learn how the presence of polymers affects the liquid crystal
smectic part of the phase diagram I combine the entropically induced
free energy in Eq.\ref{dFii} with the liquid crystal smectic free energy
$F_{lc}$ from Eq.\ref{Flc} and focus on the tilt order parameter $\vec
C$ part of the resulting free energy $F_C$
\begin{eqnarray}
F_C={1\over2}\int\bigg[&&\hspace{-.1in}J |\nabla\vec{C}|^2 + 
\big(T-T_{AC}-\tilde{n}T l_p(T)^{1/\nu_z-1}\big)|\vec{C}|^2\nonumber\\
&+&{u\over2}|\vec{C}|^4\bigg]\;.
\label{Fc}
\end{eqnarray}
By minimizing the free energy $F_C$ it is easy to see that the
SmA phase, characterized by $\langle\vec{C}\rangle=0$, is stable
in that part of the phase diagram where the coefficient
$(T-T_{AC}-\tilde{n}T l_p(T)^{1/\nu_z-1})$ of the quadratic term in
$F_C$ is positive.  The transition to the SmC phase, where
$\langle\vec{C}\rangle\neq 0$ takes place when this coefficient
changes sign and becomes negative.

For a vanishing polymer density $\tilde{n}=0$ the quadratic
coefficient is simply $T-T_{AC}$ and the SmA-SmC transition occurs at
a unique temperature $T_{AC}$, with the SmA stable at $T>T_{AC}$ and
SmC order developing for $T<T_{AC}$.  However, in the presence of
polymers ($\tilde{n}\neq 0$), depending on the $T$ dependence of
$l_p(T)$ and the polymer density $\tilde{n}$, it is possible for this
quadratic coefficient to vanish at a more than one root, corresponding
to a reentrant phase diagram for the SmA and SmC phases, of the type
displayed in Fig.\ref{fig1}. Assuming that generically $l_p(T)\sim c
T^\beta$, it is easy to see that for $\alpha\equiv\beta(1/\nu_z-1)>0$
there will be two roots, for $\tilde{n}\leq \tilde{n}^*$. For example
for $\alpha=1$ these roots can be easily found analytically
$T_{CA}^+(\tilde{n})= \big(1 + (1 - 4\tilde{n}
T_{AC})^{1/2})/(2\tilde{n}\big)$, $T_{AC}^-(\tilde{n})=
\big(1 - (1 - 4\tilde{n} T_{AC})^{1/2})/(2\tilde{n}\big)$
and determine the high- and low-$T$ parts of the phase
boundary of the reentrant transition, respectively. For a general
$\alpha$ the complete phase diagram in the $n-T$ plane is
defined by the phase boundary,
\begin{equation}
\tilde{n}(T)={(T-T_{AC})/c T^{1+\alpha}}\;
\label{boundary}
\end{equation}
which for $\alpha>0$ exhibits a reentrant behavior illustrated in
Fig.\ref{fig1}, with the SmA phase sandwiched by the SmC phase on
both the low- and high-$T$ sides. For $\alpha\leq 0$ the transition is
not reentrant but $T_{AC}(n)$ {\it increases} with the increased
polymer concentration. This latter scenario is an entropic analog of
the compression induced SmA-SmC transition discovered
experimentally over 20 years ago by Ribota, et al.\cite{Ribota}.

As expected for $\alpha>0$ the lower transition $T^-(n)$ approaches
$T_{AC}$ of a pure liquid crystal as $n\rightarrow 0$, and near
$T_{AC}$ the phase boundary is a linearly-increasing function of
$n$. The transition temperature $T_{CA}^+\rightarrow\infty$, for a
vanishing polymer density, however this behavior gets interrupted at
$\tilde{n}=\tilde{n}_c\equiv (T_{NA}-T_{AC})/c T_{NA}^{1+\alpha}$ and
$T_{CA}^+(\tilde{n}_c)\approx T_{NA}$, where the N--SmA phase boundary
gets encountered (see Fig.\ref{fig1}). For $\tilde{n}<\tilde{n}_c$ the
transition goes directly from the nematic to SmA, followed by the
standard SmA-SmC transition. From Eq.\ref{boundary} and
Fig.\ref{fig1}, it is clear that the reentrant transition, together
with the SmA phase disappears for $\tilde{n}>n^*\equiv
\alpha^\alpha/\left((\alpha+1)^{\alpha+1}c T_{AC}^\alpha\right)$.

It is useful at this stage to examine a simple model of a polymer in
order to extract the $T$ dependence of $l_p(T)$ and the
resulting prediction for the shape of the phase boundary. At short
scales a polymer exhibits an energetic rigidity against bending with a
bending modulus $\kappa$, and can be described by an effective free energy
$F_1\approx{\kappa\over2}\int d s (\partial\theta/\partial s)^2
\approx{\kappa\over2}\int d s (\partial^2\vec{r}/\partial s^2)^2$.
In a length $L$ smaller than an orientational persistent length
$l_{p1}(T)$, the polymer will explore an angular range
$\theta_{rms}\approx L T/\kappa$, corresponding to transverse
deviations $R_G\approx\theta_{rms} L\sim\sqrt{T/\kappa} L^{3/2}$. This
implies that $l_{p1}(T)$, length up to which above model is valid and
the polymer behaves as a directed one, is
$l_{p1}(T)=\mbox{Min}(\kappa/T,a_0)$, where $a_0$ is the minimum
length set by the inter-monomer distance. For this range of length
scales ($<l_{p1}(T)$), assuming $\kappa/T>a_0$, I therefore find
$\beta=-1$, $\nu_z=3/2$, $\alpha=1/3>0$, implying a reentrant
SmC--SmA--SmC transition.

In thermotropic smectics the interlayer spacing $d$ is a microscopic
length scale on the order of few angstroms and therefore a typical
length $L_0$ of a section of a confined polymer between the collisions
with smectic layers will be smaller than $l_{p1}(T)$, implying the
reentrant behavior illustrated in Fig.\ref{fig1}. For the lyotropic
smectics, however, where the interlayer spacing can be significantly
larger, the polymer collision length $L_0$ will typically be larger
than $l_{p1}(T)$. For this $L_0>>l_{p1}$ regime a more appropriate
model is that of a coiled polymer with an entropically generated
elastic modulus $\sigma(T)\approx T/l_{p1}(T)$. In this regime the
effective free energy describing polymer conformation is
$F_2\approx{\kappa\over2}\int d s (\partial\vec{r}/\partial s)^2$ and
leads to
$R_G\sim\sqrt{T/\sigma(T)}L^{1/2}\sim\sqrt{l_{p1}(T)}L^{1/2}$.
Assuming $l_{p1}=\kappa/T>a_0$, for this regime I find $\beta=-1$,
$\nu_z=1/2$, $\alpha=-1>0$. At high $T$ and for low-rigidity
polymers with small $T$-independent persistent length
$l_{p1}=a_0$ $\beta=0$, $\nu_z=1/2$, and $\alpha=0$. Both cases imply
a single non-reentrant SmA--SmC transition, with $T_{AC}(n)$
monotonically shifting to higher $T$ with increasing polymer
concentration.

When additional thermal fluctuations of the $\vec C$ order parameter
are taken into account, as usual, the mean-field exponents and the
actual shape of the phase boundary will be quantitatively
modified. Based on general symmetry arguments, I expect that both the
high- and low-$T$ transitions will be in the same universality
class, that of the 3d XY model, which describes the
conventional (polymer-free) SmA-SmC
transition\cite{deGennes}. It is unfortunately difficult to make
precise, model independent quantitative predictions of how these
additional fluctuations will modify the reentrant phase boundary,
except to argue that they will tend to partly wash it
out. Nevertheless it is likely that even in the presence of these
additional fluctuations, the reentrant behavior and the upward
$T_{AC}(n)$ shift with increasing $n$, discussed here, should still be
observable in thermotropic and lyotropic liquid crystals,
respectively.

A more important obstacle to observing the reentrant phase diagram
proposed here lies in the Eq.\ref{boundary}. Maximizing $n(T)$ with
respect to $T$, one can easily show that the turning point of this
phase boundary (responsible for reentrant behavior) occurs at
$T^*=T_{AC}(\alpha+1)/\alpha$, which is unfortunately too high of a
temperature to be experimentally observable in common liquid crystal
systems with $T_{AC}\approx 300 K$. At such high $T$, the entropic
effects discussed here will be dominated by other thermal effects such
as for example the transition into the isotropic phase and 3d
deconfinement of polymers. However, as discussed above, it is likely
that the shape and therefore the $T_{AC}(\alpha+1)/\alpha$ estimate
for the maximum in the phase boundary will be modified by
fluctuations, possibly allowing the reentrance to be observable in
some liquid crystal systems.

Finally, one additional scenario is possible. As argued in this paper,
the reentrance mechanism is driven by the fact that at high
$T$ the tilt in the SmC phase is more favorable for
accommodating the out-of-plane undulation of the confined polymers,
thereby lowering their free energy. However, for this accommodation to
take place it is not necessary to have long-range order in the
SmC order parameter; all that is necessary is tilt, without
$\vec{C}$ directional long-range correlations. Therefore it is
possible that the high $T$ part of the SmC phase in
Fig.\ref{fig1} is replaced by the SmA phase with an enhanced
local $\vec{C}$ order, $<|\vec{C}|>\neq 0$, but with $<\vec{C}>= 0$
due to random $\vec{C}$ orientations. In this case the high
temperature $T_{CA}$ transition will be replaced by a crossover or a
first order transition within SmA phase.

In summary, I have argued that the smectic part of a liquid crystal
phase diagram can be considerably modified by the presence of
polymers, confined between the smectic layers. I have shown that this
confinement generates an effective entropic interaction which can lead
to a reentrant SmC--SmA--SmC phase diagram.

I thank J. Toner, N. Clark and V. Ginzburg for discussions and
constructive criticism, and acknowledge financial supported by the NSF
Grant DMR-9625111.

\end{multicols}
\end{document}